\documentclass[%
reprint,
amsmath,amssymb,
floatfix,longbibliography
]{revtex4-1}
\usepackage{graphicx}
\usepackage{dcolumn}
\usepackage{mathrsfs}
\usepackage{bm}
\usepackage{hyperref}

\usepackage{xcolor}
\usepackage[normalem]{ulem}

\newcommand{\B}{\hat{b}}
\newcommand{\n}{\hat{n}}

\begin{document}
	\title{Mean-field study of the Bose-Hubbard model in Penrose lattice}
	\author{Rasoul Ghadimi, Takanori Sugimoto, Takami Tohyama}
	\affiliation{Department of Applied Physics, Tokyo University of Science, Tokyo 125-8585, Japan}
	\date{\today}
	\begin{abstract}
		We examine the Bose-Hubbard model in the Penrose lattice based on inhomogeneous mean-field theory. Since averaged coordination number in the Penrose lattice is four, mean-field phase diagram consisting of the Mott insulator (MI) and superfluid (SF) phase is similar to that of the square lattice. However, the spatial distribution of Bose condensate in the SF phase is significantly different from uniform distribution in the square lattice. We find a fractal structure in its distribution near the MI-SF phase boundary. The emergence of the fractal structure is a consequence of cooperative effect between quasiperiodicity in the Penrose lattice and criticality at the phase transition.
	\end{abstract}
	\maketitle
	
	\section{Introduction}
	\label{introduction}
	Quasicrystals have aperiodic structure different from fully disordered one. Although translational symmetry is absent, the presence of sharp spots in Bragg reflection indicates long-range order~\cite{PhysRevLett.53.1951,PhysRevLett.53.2477}. Quasicrystals can be realized even in bilayer graphene~\cite{Yao6928} and photonic lattices~\cite{doi:10.1063/1.4754136}. In addition to various characteristics due to aperiodicity~\cite{Steurer:ib5056,doi:10.1146annurev.matsci.38.060407.130318}, recent new findings expand the field of quasicrystal to include superconductivity~\cite{Kamiya2018,PhysRevB.100.014510,PhysRevB.95.024509,PhysRevResearch.1.022002}, quantum criticality~\cite{Deguchi2012,doi:10.7566/JPSJ.85.073712}, and topology~\cite{PhysRevLett.108.220401,doi:10.7566/JPSJ.83.083707,PhysRevLett.119.215304,PhysRevB.100.081405,PhysRevLett.121.126401}.  In general, self-similarity in quasicrystals dictates fractal structure in wavefunction and phase diagram ~\cite{Rasoul2017,PhysRevX.6.011016}.  This characteristic is justified by the presence of the inflation and deflation rules to construct quasicrystals~\cite{PhysRevB.39.9904}.   
	
	One of the well-known two-dimensional (2D) quasicrystals is the so-called Penrose lattice~\cite{DEBRUIJN198139,DEBRUIJN198153}. One can construct the lattice using inflation, projection, or multi-grade rules. The Penrose lattice has been studied intensively~\cite{oitmaa1990antiferromagnetic,PhysRevB.43.1378,PhysRevB.77.104427,PhysRevB.93.075141,Takemori_2015,PhysRevB.96.214402,takemori2018intersite,Jagannathan2012,PhysRevB.79.172406,PhysRevB.101.195118} and its structure dictates thermodynamically degenerate states in energy spectrum~\cite{PhysRevB.38.1621,PhysRevB.37.2797}.  
	
	Ultracold gases in optical lattices provide us an ideal playground of strong correlation~\cite{Bloch2005} and also quasicrystals~\cite{PhysRevA.92.063426,PhysRevLett.79.3363,PhysRevA.72.053607,Corcovilos:19,PhysRevLett.120.060407,PhysRevLett.111.185304}, which allows us to investigate the interplay of strong correlation and aperiodicity. A typical strongly correlated system in optical lattice is the Bose-Hubbard model, where phase transition between Mott insulator (MI) to superfluid (SF) phase appears~\cite{PhysRevB.40.546,Dutta_2015} as experimentally observed~\cite{PhysRevLett.81.3108,Greiner2002}. 
	The Bose-Hubbard model is also used to describe the effective low-energy theory of superconducting films and arrays of Josephson junctions \cite{Giamarchi2008,PhysRevLett.84.5868,PhysRevB.47.342}. 
	Recent achievements in establishing an eight-fold rotationally symmetric optical lattice attract new attention \cite{PhysRevLett.122.110404}, in connection with theoretical investigation of an extended Bose-Hubbard with quasicrystalline confined potential~\cite{PhysRevA.100.053609}, where spontaneous breaking of underlying eight-fold symmetry is observed. However, the effect of aperiodicity in the Bose-Hubbard model is not yet fully understood both theoretically and experimentally.

	\begin{figure}[h!]
		\centering
		\includegraphics[width=0.8\linewidth]{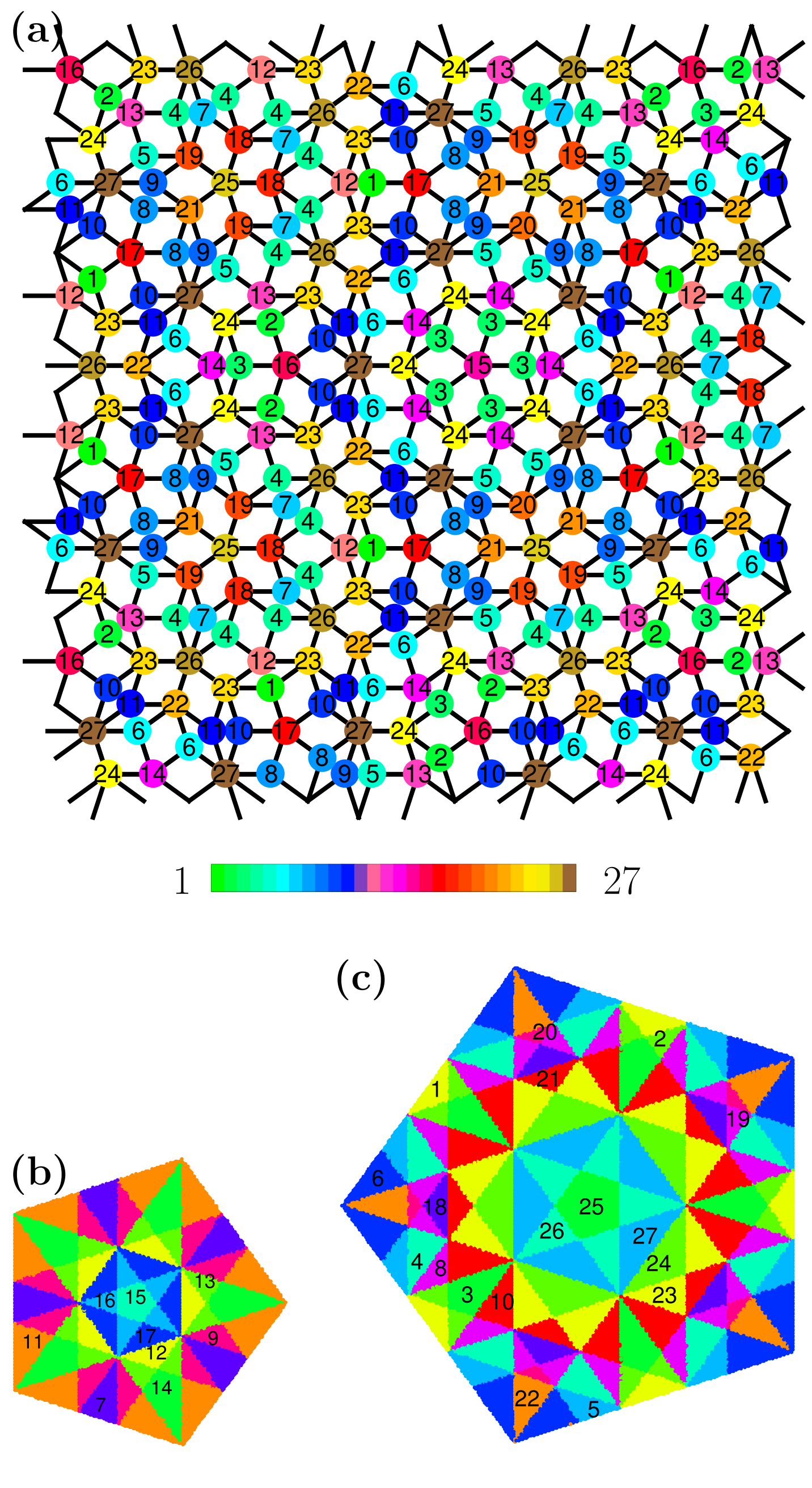}
		\caption{(a) Part of Penrose lattice. The number and color in each vertex indicate the index $\alpha$ of vertices among 27 different kinds of vertices. (b) Perpendicular space of Penrose lattice for $\mathcal{Z}=1$, and (c) that for $\mathcal{Z}=2$. The number is the same as (a). Different colors in (b) and (c) distinguish different sections in perpendicular space. }
		\label{fig:crystall}
	\end{figure}
	
	\begin{table*}[th!]
		\caption{Link configuration of distinct vertices in Penrose lattice. Listed are index $\alpha$ determined in the present work, the total number of paths using $k$ links, $M_k$ ($k=1,2,3$), the number of vertices having $l$ links, to which one can access using $k$ links, $m_k^{(l)}$ ($l=3,4,5,6, 7$). Note that $\sum_l m_k^{(l)}=M_k$.}
		\begin{ruledtabular}
			\begin{tabular}{cccccccccccccccccccccccccccc}
				$\alpha$ & 1 & 2 & 3 & 4 & 5 & 6 & 7 & 8 & 9 & 10 & 11 & 12 & 13 & 14 & 15 & 16 & 17 & 18 & 19 & 20 & 21 & 22 & 23 & 24 & 25 & 26 & 27 \\ \hline
				$M_1$ & 3 & 3 & 3 & 3 & 3 & 3 & 3 & 3 & 3 & 3 & 3 & 4 & 4 & 4 & 5 & 5 & 5 & 5 & 5 & 5 & 5 & 5 & 5 & 5 & 5 & 6 & 7 \\
				$m_1^{(3)}$ & 0 & 0 & 0 & 0 & 0 & 0 & 0 & 0 & 0 & 0 & 0 & 2 & 2 & 2 & 5 & 5 & 5 & 4 & 4 & 4 & 4 & 4 & 3 & 2 & 0 & 3 & 6 \\
				$m_1^{(4)}$ & 0 & 0 & 0 & 1 & 1 & 1 & 0 & 0 & 0 & 0 & 0 & 0 & 0 & 0 & 0 & 0 & 0 & 0 & 0 & 0 & 0 & 0 & 1 & 2 & 0 & 0 & 0 \\
				$m_1^{(5)}$ & 3 & 3 & 3 & 1 & 1 & 1 & 2 & 2 & 2 & 2 & 2 & 2 & 2 & 2 & 0 & 0 & 0 & 1 & 1 & 1 & 1 & 0 & 0 & 0 & 5 & 3 & 1 \\
				$m_1^{(6)}$ & 0 & 0 & 0 & 1 & 0 & 0 & 1 & 0 & 0 & 0 & 0 & 0 & 0 & 0 & 0 & 0 & 0 & 0 & 0 & 0 & 0 & 1 & 1 & 0 & 0 & 0 & 0 \\
				$m_1^{(7)}$ & 0 & 0 & 0 & 0 & 1 & 1 & 0 & 1 & 1 & 1 & 1 & 0 & 0 & 0 & 0 & 0 & 0 & 0 & 0 & 0 & 0 & 0 & 0 & 1 & 0 & 0 & 0 \\
				$M_2$ & 15 & 15 & 15 & 15 & 16 & 16 & 16 & 17 & 17 & 17 & 17 & 16 & 16 & 16 & 15 & 15 & 15 & 17 & 17 & 17 & 17 & 18 & 19 & 21 & 25 & 24 & 23 \\
				$m_2^{(3)}$ & 11 & 10 & 9 & 9 & 12 & 12 & 11 & 15 & 14 & 14 & 13 & 6 & 5 & 4 & 0 & 0 & 0 & 0 & 0 & 0 & 0 & 3 & 5 & 10 & 20 & 10 & 2 \\
				$m_2^{(4)}$ & 2 & 3 & 4 & 0 & 0 & 0 & 0 & 0 & 0 & 1 & 1 & 4 & 5 & 6 & 0 & 0 & 0 & 2 & 2 & 2 & 0 & 2 & 0 & 0 & 0 & 4 & 4 \\
				$m_2^{(5)}$ & 0 & 0 & 0 & 6 & 4 & 3 & 5 & 2 & 3 & 1 & 1 & 2 & 2 & 2 & 15 & 13 & 11 & 11 & 11 & 11 & 13 & 9 & 12 & 11 & 5 & 4 & 10 \\
				$m_2^{(6)}$ & 2 & 1 & 0 & 0 & 0 & 1 & 0 & 0 & 0 & 1 & 2 & 4 & 2 & 0 & 0 & 0 & 0 & 4 & 2 & 0 & 0 & 0 & 0 & 0 & 0 & 6 & 0 \\
				$m_2^{(7)}$ & 0 & 1 & 2 & 0 & 0 & 0 & 0 & 0 & 0 & 0 & 0 & 0 & 2 & 4 & 0 & 2 & 4 & 0 & 2 & 4 & 4 & 4 & 2 & 0 & 0 & 0 & 7 \\
				$M_3$ & 53 & 55 & 57 & 57 & 56 & 57 & 58 & 55 & 57 & 57 & 60 & 68 & 71 & 74 & 75 & 79 & 83 & 87 & 89 & 91 & 93 & 90 & 89 & 85 & 85 & 102 & 121 \\
			\end{tabular}
		\end{ruledtabular}	
		\label{tab}
	\end{table*}
	
	In this paper, we investigate the phase diagram of the Bose-Hubbard model in the Penrose lattice. We use a self-consistent mean-field theory and find that the distribution of Bose condensate in the Penrose lattice exhibits a fractal structure near the MI-SF boundary. We attribute the appearance of the fractal structure to a consequence of the divergence of correlation length seen in any phase transition. Therefore, the fractal structure is a common signature of phase transition in quasiperiodic systems. 
	
	The arrangement of this paper is as follows.  In Sec.~\ref{model}, we describe our Bose-Hubbard model on the Penrose lattice and mean-field treatment. The classification of lattice sites (vertices) is also introduced. In Sec.~\ref{results}, we discuss the result of phase diagram, local superfluid amplitude, and a critical behavior of several quantities. A fractal structure near the phase transition in the perpendicular space in the Penrose lattice is also discussed. Finally, a summary is given in Sec.~\ref{conclusion}.

	\section{Model and method}	
	\label{model}
	The Hamiltonian of the single-band Bose-Hubbard model is defined by
	\begin{equation}\label{Bose-Hubbard}
	H_{BH}=-J\sum_{<i,j>}(\B^{\dagger}_i\B_j+\B^{\dagger}_j\B_i)-\mu\sum_i \n_i+\frac{U}{2}\sum_i \n_i(\n_i-1),
	\end{equation}
	where $\B_i$ and $\B^{\dagger}_i$ are annihilation and creation operators of bosons at site $i$ and the number operator $\n_i=\B^{\dagger}_i\B_i$. We refer the site to vertex, which is denoted by circles in Fig.~\ref{fig:crystall}(a). The summation $\langle i,j\rangle$ represents nearest-neighbor (NN) links in the Penrose lattice shown as short bar connecting two vertices in Fig.~\ref{fig:crystall}(a). $J$, $\mu$, and $U$ in Eq.~(\ref{Bose-Hubbard}) are the hopping energy of boson, the chemical potential, and on-site Coulomb interaction, respectively.
	We note that hopping processes with the shortest inter-vertex distance, for example, hopping between numbers 8 and 9 in Fig.~\ref{fig:crystall}(a), are not included in Eq.(\ref{Bose-Hubbard}). This exclusion guarantees bipartite properties of this Penrose lattice.
	
	Because of the presence of the hopping term in Eq.~(\ref{Bose-Hubbard}), the exact solution is inaccessible. Therefore, we use a mean-field technique and decouple the hopping term using local superfluid amplitude $\langle \B_i\rangle$. The resulting mean-field Hamiltonian is given by $H_\mathrm{MF}=\sum_i H_i+E_0$ with
	\begin{equation}\label{IMFT}
	H_i=-J\left( \psi_i^* \B_i+H.c.\right)-\mu \n_i+\frac{U}{2} \n_i(\n_i-1),
	\end{equation}
	where $\psi_i=\sum_{j\in \text{NN.i}}\langle\B_j\rangle$ with summation over NN links connected to the vertex $i$ and $E_0=J\sum\psi^*_i\langle \B_i\rangle$.
	
	In order to obtain a self-consistent solution of Eq.~(\ref{IMFT}) in the local Hilbert space containing maximally $n_\mathrm{b}$ bosons, we start with an initial $\psi_i$ and then calculate $\langle \n_i\rangle$ and $\langle \B_i\rangle$ using the ground-state wavefunction for each vertex. We continue updating $\psi_i$ until the convergence of $\langle \n_i\rangle$ and $\langle \B_i\rangle$ is obtained within a certain tolerance ($10^{-9}$ in our case). 
	This self-consistent procedure gives rise to site-dependent distribution of $\langle \n_i\rangle$ and $\langle \B_i\rangle$ on the Penrose lattice. This technique is sometimes called inhomogeneous mean-field theory~\cite{PhysRevB.85.214524,doi:10.1002/andp.201100274}, which gives equivalent results with variational Gutzwiiler method~\cite{PhysRevA.83.053608,PhysRevA.86.013623,Barman_2014,PhysRevLett.98.260405,PhysRevLett.75.4075,PhysRevA.91.043632,PhysRevB.44.10328,PhysRevB.45.3137}. We note that this self-consistent procedure gives moderately consistent results compared by quantum Monte Carlo simulations in determining the phase diagram of the Bose-Hubbard model with NN repulsion interaction~\cite{PhysRevB.86.054520}.
	
	We take $n_\mathrm{b}=7$. Within our mean-field theory, we generally find the MI and SF phases in the Bose-Hubbard model.  In the MI phase, all sites have equal integer number of  bosons and thus $\langle \B_i\rangle=0$. On the other hand, $\langle \B_i\rangle$ is nonzero for the SF phase. In our method, we find order parameters on all vertices. Therefore, we can check the existence of exotic states like Bose-glass, supersolid, and density-wave phases. We did not see these phases in our model. This is reasonable since these phases appear in the presence of disorder and/or NN interaction in Eq.~(\ref{Bose-Hubbard})~\cite{PhysRevA.83.051606,doi:10.1002/andp.201100274,PhysRevLett.75.4075}. To minimize the boundary effects, we apply periodic boundary conditions (PBC) in an approximant of Penrose lattice containing  $N=167761$ vertices (For detail, see the Appendix.~\ref{PBC})~\cite{entin1988penrose,BABALIEVSKI199027,PhysRevB.43.8879}. 
	
	In the Penrose lattice, we can classify vertices in terms of their local environment. For this classification, we first find the number of NN links, $M_1$, i.e., the total number of paths using one link ($M_k$ with $k=1$ and see the second row in Table~\ref{tab}), which is equivalent to coordination number for each vertex in Fig.~1(a). $M_1$ changes from $3$ to $7$. This means that all of sites are indexed by five kinds of vertices. Next, we count the number of NN vertices having $l$ links, $m_1^{(l)}$, and make a list of them (the third-seventh rows in Table~\ref{tab}). From the list of $m_1^{(l)}$ together with $M_1$, we find fourteen types of configurations, meaning that all of sites are indexed by fourteen kinds of vertices. The total number of paths using two links ($k=2$) from a given vertex is then expressed as $M_2=\sum_{l=3}^7 m_1^{(l)} l$, which is listed in the eighth row of Table~\ref{tab}. We repeat this listing for the vertices accessed by using the two links form a given vertex, which is shown in the ninth-thirteenth rows as $m_2^{(l)}$. In the last row of Table~\ref{tab}, the total number of paths using three links ($k=3$) from a given vertex ($M_3=\sum_{l=3}^7 m_2^{(l)} l$) is listed. Performing this procedure for all vertices in our supercell, we find that there are twenty-seven kinds of vertices, by which almost the whole system is covered. They are indexed as $\alpha (=1,2,\cdots, 27)$ in the first row of Table~\ref{tab}.  We note that, in our approximant periodic lattice with $N=167761$ vertices, there are vertices that do not belong to the 27 types around defects, but we can ignore them since the total number of the defects is just 2.
	
	We call the number of distinct vertices for a given $k$ the number of classes (NoC).  For example, NoC is equal to 5, 14, and 27 for $k=1$, 2, and 3, respectively.  We draw a small portion of Penrose lattice in Fig.~\ref{fig:crystall}(a), where each vertex has an index $\alpha (=1,2,\cdots, 27)$ and color indicating its class obtained for $k=3$. We can increase $k$ as many as possible.  We find NoC $\propto k^{1.84}$ in the large $k$ region (see Fig.~\ref{fig:MeanFieldComparison}(c)). We will come back to this point later. 
	

	As explained in the Appendix.~\ref{PBC}, vertexes in the Penrose lattice can be labeled with five integers, originated from cut and projection of five dimensional cubic lattice~\cite{PhysRevB.96.214402}. One can construct original Penrose lattice by mapping those labels. However, using another mapping, one finds four different 2D structures, called perpendicular space, where we assign the four structures to $\mathcal{Z}=1$, 2, 3, and 4 [see Figs.~\ref{fig:crystall}(b) and \ref{fig:crystall}(c) for $\mathcal{Z}=1$ and 2, respectively]. We can divide perpendicular space into symmetric sections, where each section represents vertex with similar local circumstances. Therefore, one notices the index $\alpha$ in Fig.~\ref{fig:crystall}(a) mapped to different sections in the perpendicular space [see Figs.~\ref{fig:crystall}(b) and \ref{fig:crystall}(c)]. We note that the bipartite property of Penrose lattice leads to the fact that $\mathcal{Z}=1,3$ and $\mathcal{Z}=2,4$ belong to different subsystems, though the same $\alpha$ are shared among them.

	\section{Results}
	\label{results}
\begin{figure}[t]
		\centering
		\includegraphics[width=0.9\linewidth]{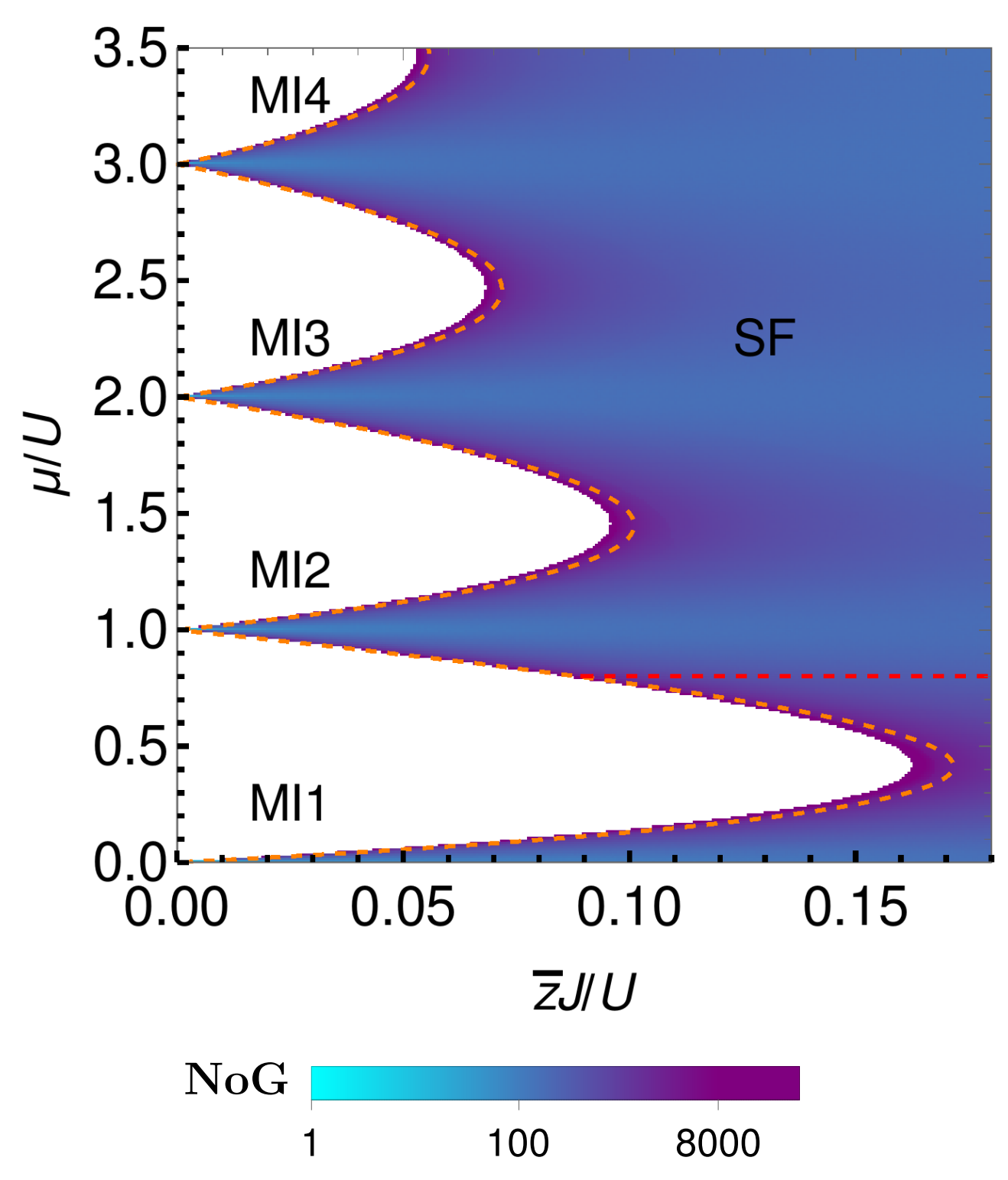}
		\caption{Phase diagram of the Bose-Hubbard model on the Penrose lattice. The white area with the shape of lobes corresponds to the MI phase with $n_0$ bosons in all vertices, denoted by MI$n_0$ ($n_0=1,2,\cdots$). In the SF phase, the number of gap (NoG) for a threshold value of $10^{-7}$ defined in the text is plotted with color scale. The analytical MI-SF mean-field phase boundary for the square lattice in Eq.~(\ref{Eq:SquareLattice}) is plotted by the orange dashed curve.}
		\label{fig:phaseDiagram}
	\end{figure}

	We first examine the phase diagram of the Bose-Hubbard model on the Penrose lattice. From the calculation of two order parameters per vertex, $\langle  \n_i\rangle$ and $\langle \B_i\rangle$, 
	for the Bose-Hubbard model without disorder and/or inter-site interaction,
	we expect two phases: one is MI with $\langle \B_i\rangle=0$ and $\langle  \n_i\rangle=n_0$ ($n_0=1,2,\cdots$, corresponding to bosonic occupation number at each vertex), and the other is SF with $\langle \B_i\rangle\neq 0$. 
	In fact, we find none of Bose-glass,  density wave, and supersolid phases in the phase diagram. 
	Figure~\ref{fig:phaseDiagram} shows the phase diagram, where we find MI phases denoted by MI$n_0$ and SF. Since averaged coordination number in the Penrose lattice is $\bar{z}=4$, which is the same as the coordination number $z=4$ in the square lattice, the phase boundary between MI and SF is expected to be similar to that of the square lattice. This is the case as shown by the dashed orange curve along MI lobes in Fig.~\ref{fig:phaseDiagram}, which is mean-field phase boundary for the square lattice given analytically~\cite{PhysRevA.63.053601,Freericks_1994} by
	\begin{equation}\label{Eq:SquareLattice}
	zJ_{c}/U=\frac{-\frac{\mu}{U}-(\frac{\mu}{U})^2+s+2\frac{\mu}{U} s-s^2}{1+\frac{\mu}{U}},
	\end{equation}
	where $s=\text{round}(\mu/U+1/2)$. The similarity indicates small effect of aperiodicity on the phase boundary.
	
	In the SF phase of square lattice, the local superfluid amplitude $\langle \B_i\rangle$ is uniform, i.e., independent of $i$, for any region in the phase diagram. On the other hand, nonuniform distribution of $\langle \B_i\rangle$ in the Penrose lattice is easily expected from the presence of different types of vertices as shown in Fig.~\ref{fig:crystall}(a). Then, an arising question is how its nonuniform distribution changes in the phase diagram. To see this, we define an $\alpha$ dependent average of $\langle \B_i\rangle$ as $\overline{b}_\alpha=N_\alpha^{-1}\sum_{i\in\alpha}\langle \B_i\rangle$, where $N_\alpha$ is the number of $\alpha$-type vertex in the whole lattice. This quantity can distinguish the twenty-seven classes of vertices. However, each class should have further internal structure coming from possible extension of NoC for $k\ge 4$. To recognize this structure, we also define a mean deviation of local superfluid amplitude distribution as $\delta b_\alpha=\sqrt{N_\alpha^{-1}\sum_{i\in\alpha}(\langle \B_i\rangle-\overline{b}_\alpha)^2}$. 
	
	\begin{figure}[t]
		\centering
		\includegraphics[width=\linewidth]{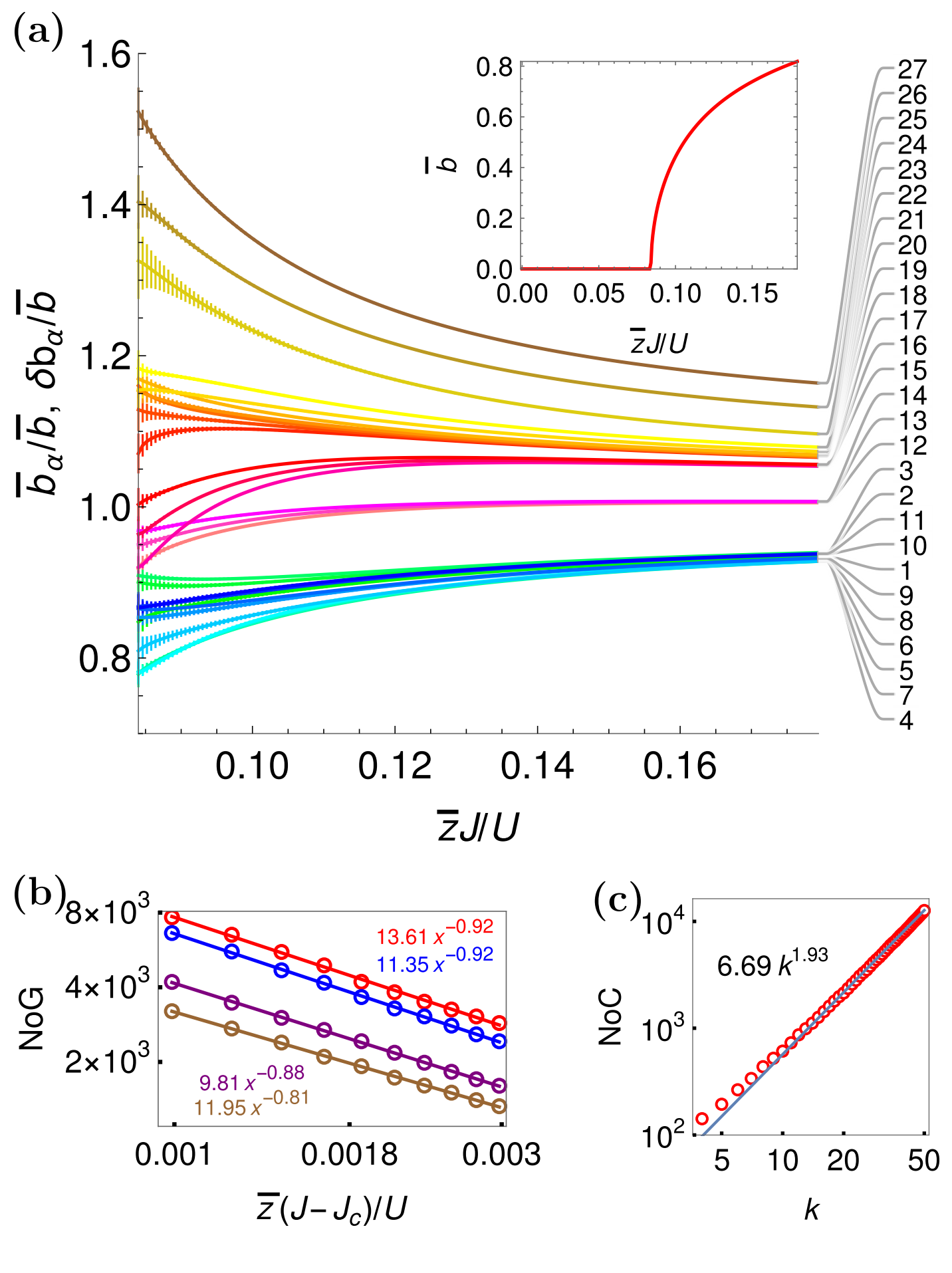}
		\caption{
			(a) Averaged order parameters $\overline{b}_\alpha/\overline{b}$ and mean deviation $\delta b_\alpha/\overline{b}$ as a function of $\bar{z}J/U$ along the horizontal dotted line ($\mu/U=0.8$) in Fig.~\ref{fig:phaseDiagram}. The colored curves represent $\overline{b}_\alpha/\overline{b}$. The bars centered at each curve represent $\delta b_\alpha/\overline{b}$. The color scheme is the same as Fig.~\ref{fig:crystall}(a) and the number denoted at the right-hand side indicates each class $\alpha$ as shown in Fig.~\ref{fig:crystall}(a). Inset shows averaged order parameter $\overline{b}$.
			(b) Log-log plot of the number of gaps (NoG) defined in the text as a function of $\bar{z}(J-J_c)/U$ along the horizontal dotted line in Fig.~\ref{fig:phaseDiagram}. The brown, purple, blue and red circles represent NoG for threshold values of $10^{-5}$, $5\times10^{-6}$, $10^{-6}$, $5\times10^{-7}$, respectively. The lines represent fitting function denoted by the corresponding color, where $x=\bar{z}(J-J_c)/U$. 
			(c) Log-log plot of the number of classes (NoC) as a function of the number of links $k$. The blue line represents a fitting function shown in the figure.}
		\label{fig:MeanFieldComparison}
	\end{figure}
	
	With approaching phase transition from the superfluid side, the average value of local superfluid amplitude, $\overline{b}=\sum_i\langle \B_i\rangle/N$ with $N=\sum_\alpha N_\alpha$, reduces its value toward zero as shown in the inset of Fig.~\ref{fig:MeanFieldComparison}(a). At the same time, both $\overline{b}_\alpha$ and $\delta\overline{b}_\alpha$ become smaller. Therefore, we use $\delta\overline{b}_\alpha/\overline{b}$ to evaluate the magnitude of the mean deviation of $\delta\overline{b}_\alpha$.  Note that the larger $\delta b_{\alpha}/\overline{b}$ is, the deeper the internal structure is.
	
	In Fig.~\ref{fig:MeanFieldComparison}(a), we plot $\overline{b}_\alpha/\overline{b}$ and $\delta b_\alpha/\overline{b}$ as a function of $\bar{z}J/U$ along the horizontal dotted line ($\mu/U=0.8$) in Fig.~\ref{fig:phaseDiagram}.
	We note that $\delta b_\alpha/\overline{b}$ is denoted by the length of bars for each $\overline{b}_\alpha/\overline{b}$. At large $\bar{z}J/U$ far from the phase boundary, $\overline{b}_\alpha/\overline{b}$ is tend to be grouped accompanied by negligibly small $\delta b_\alpha/\overline{b}$. In the limit of $\bar{z}J/U\rightarrow\infty$, $\overline{b}_\alpha/\overline{b}$ is grouped into five classes equivalent to the coordination number, i.e., NoC for $k=1$. This means that, if correlation effect is small, the coordination number controls physical properties as expected. On the other hand, with approaching $\bar{z}J/U$ to the phase boundary, the mean deviation $\delta b_\alpha/\overline{b}$ becomes large. This means that the number of distinct vertices with different local superfluid amplitude increases with approaching to the boundary. In other words, long-distant correlation becomes important in order to obtain critical behaviors near the phase transition.
	
	\begin{figure*}
		\centering
		\includegraphics[width=0.93\linewidth]{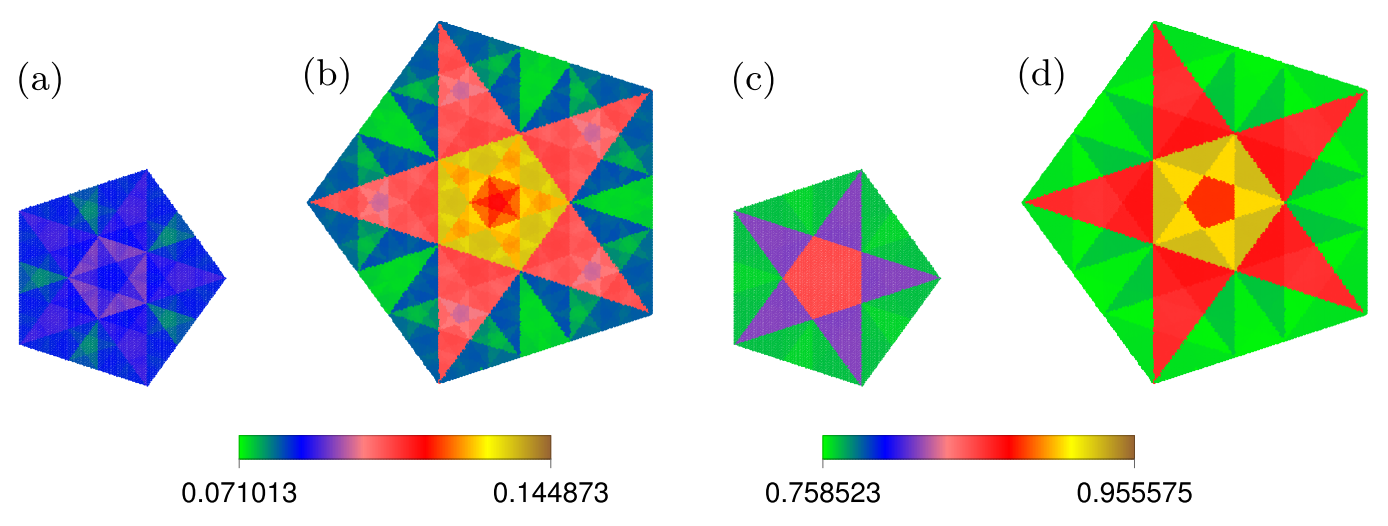}
		\caption{Representation of vertex  local superfluid amplitude in perpendicular space ($\mathcal{Z}=1,2$) for (a,b) $(\mu/U,\bar{z}J/U)=(0.8,0.084)$ and for (c,d) $(\mu/U,\bar{z}J/U)=(0.8,0.18)$. The numbers in color bars show extremes of local superfluid amplitude for given parameters.  }
		\label{fig:PrependicularSpace}
	\end{figure*}
	
	In order to make critical behaviors visible, we introduce a new quantity that can characterize distinct number of vertices more than twenty seven. We use $\langle \B_i\rangle$ itself for this purpose, and try to find how many distinct values exist with approaching to the phase boundary. For distinguishing different value of local superfluid amplitude, we 
	i) make shifting and scaling for $\langle \B_i\rangle$ to be located within $[0,1]$. 
	This is done by evaluating $\left(\langle \B_i\rangle-\text{min}[{\{\langle \B_i\rangle\}}]\right)/\left(\text{max}[{\{\langle \B_i\rangle\}}]-\text{min}[{\{\langle \B_i\rangle\}}]\right)$, where min and max denote minimum and maximum among all values of local superfluid amplitude, respectively. Then, we
	ii) sort the scaled $\langle \B_i\rangle$ from 0 to 1, iii) calculate the difference of $\langle \B_i\rangle$ between $i$ and $i+1$ from $i=1$ to $i=N-1$, and iv) count the number of the difference (gap) whose magnitude is more than a given small threshold value. We call this number the number of gap (NoG). For example,  NoG is zero for the square lattice because $\langle \B_i\rangle$ is independent of $i$. In the Penrose lattice, we have NoG$=4$ in the large limit of $\bar{z}J/U$ since there are five distinct values of $\langle \B_i\rangle$. We show log-log plot of NoG in Fig.~\ref{fig:MeanFieldComparison}(b) along the horizontal dotted line in Fig.~\ref{fig:phaseDiagram}, where four different threshold values, $10^{-5}$ , $5\times10^{-6}$, $10^{-6}$, and $5\times10^{-7}$ are used. With approaching to the phase boundary at $\bar{z}J_c/U\approx0.0835$, NoG increases, indicating the increase of distinct vertices 
	with different local superfluid amplitude. 
	Interesting is that, with decreasing the threshold value, NoG rapidly increases near the boundary and shows a diverging behavior with an approximate exponent around $-0.9$, i.e., NoG $\propto  (J-J_c)^{-0.9}$. This resembles to a critical behavior toward continuous phase transition as suggested from the vanishing of averaged order parameter $\bar{b}$ [see inset of Fig~\ref{fig:MeanFieldComparison}(a)]. 
	
	In order to understand this diverging behavior more, we focus on the fact that the increase of NoG corresponds to the increase of distinct vertices with different local superfluid amplitude. The latter is measured by NoC, whose large region is proportional to $k^{1.84}$ as shown in Fig.~\ref{fig:MeanFieldComparison}(c). Therefore, diverging behavior in NoG is directly connected to diverging behavior in NoC at large $k$. Since $k$ represents the number of links from a given vertex, we may regard $k$ as a measure of correlation length $\xi$ from a given vertex. Based on this reasoning, we have NoG $\propto$ NoC $\propto k^{1.84} \propto \xi^{1.84}$. Since $\xi \propto (J-J_c)^{-0.5}$ for the mean-field phase transition, we finally expect that NoG $\propto (J-J_c)^{-0.92}$,  whose exponent is close to the calculated one in NoG, $\approx-0.9$. This indicates that diverging behavior in NoG is a consequence of criticality in the mean-field phase transition. We note that this critical behavior does not appear if $\mu/U=n_0$ and $J/U\rightarrow 0$. We suggest that the assumption of $\xi \propto k$ may change if we alter the class of universality. We leave this to future work. 
	
	
	Usefulness of perpendicular space presentation has already been found in considering magnetism on the Penrose lattice~\cite{PhysRevB.96.214402,PhysRevB.77.104427}.  Therefore, we show the perpendicular space representation of $\langle \B_i\rangle$ in Fig.~\ref{fig:PrependicularSpace} for two sets of parameters at the end of the red dashed line in Fig.~\ref{fig:phaseDiagram}. We  recognize notable differences in the two cases. For the parameter far from the phase boundary, we find fourteen distinct sections in Figs.~\ref{fig:PrependicularSpace}(c) and \ref{fig:PrependicularSpace}(d). The number corresponds to  the number of distinct vertices obtained by setting $k=2$ as discussed above.  On the other hand, for the parameter close to the phase boundary, we can see a fractal structure in Figs.~\ref{fig:PrependicularSpace}(a) and \ref{fig:PrependicularSpace}(b). For example, we find a various size of star structure inside stars. We can understand the emergence of the fractal structure near the phase transition as follows. Because of diverging behavior in NoG near the MI-SF phase boundary, all distances become relevant.  We have found from the previous discussion that tracing far distant links by increasing $k$ enhances NoC dramatically. Therefore we can expect further distinguishable sections in the perpendicular space, resulting in fractal nature. In other words, a combination of criticality leading to phase transition and aperiodicity is a key for the emergence of fractal structure.  
	
	\section{Conclusion}
	\label{conclusion}
	We have obtained mean-field phase diagram in the Penrose-Bose-Hubbard model.  We have found that the Penrose lattice does not change the MI-SF boundary drastically in comparison with square lattice.  However, the spatial distribution of Bose condensate is unequal, and indeed fractal structure appears in the perpendicular representation of local superfluid amplitude near the MI-SF phase transition. This is a consequence of the cooperative effect of criticality leading to phase transition and quasiperiodicity, which is expected to be a common feature in aperiodic strongly correlated systems.
	
	\section{Acknowledgments}
	This work was supported by Challenging Research Exploratory (Grant No. JP17K18764), Grant-in-Aid for Scientific Research on Innovative Areas (Grant No. JP19H05821). 
	
	\appendix
	\section{Periodic boundary condition and perpendicular representation in Penrose lattice}\label{PBC}
	\begin{figure}
		\centering
		\includegraphics[width=0.5\linewidth]{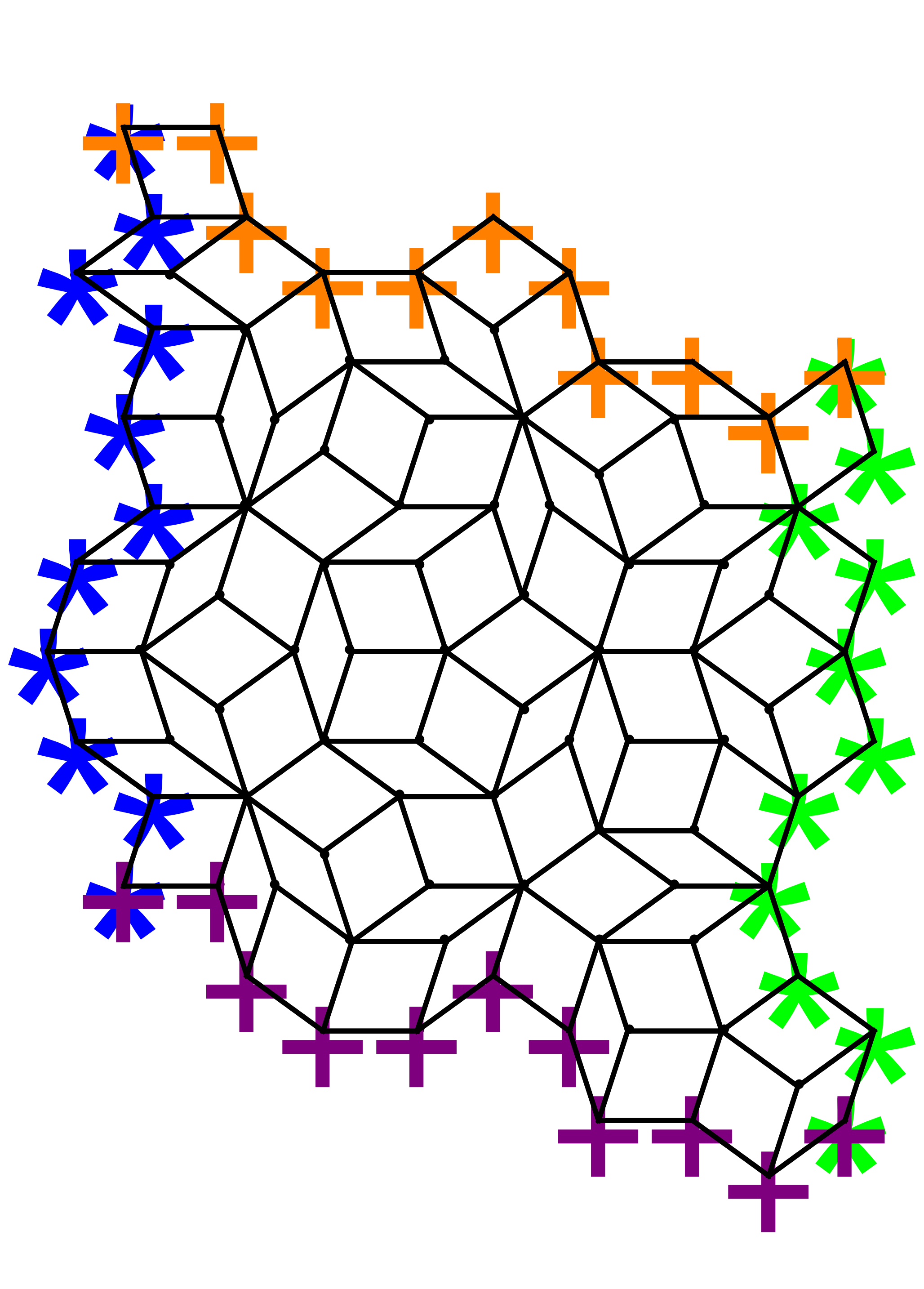}
		\caption{An approximant of Penrose lattice with an approximated golden ratio $\tau\approx\tau_3=3/2$, including $N_v=76$ vertices in a unit cell. With periodic boundary condition, orange crosses on the upper (green stars on the right) boundaries are identified with purple crosses on the lower (blue stars on the left) boundaries.}
		\label{fig:PBC}
	\end{figure}
	
	In the appendix, we explain how to construct a two-dimensional finite-size Penrose lattice with periodic boundary condition systematically.
	To make a finite-size Penrose lattice, a cut-and-projection method is well known and established.
	In this method, we first consider a hypercubic lattice in a five-dimensional space consisting of two-dimensional real space and three-dimensional perpendicular space.
	Five primitive vectors of the hypercubic lattice are given by,
	\begin{equation}
	\bm{e}_i = (\delta_{i,1},\delta_{i,2},\delta_{i,3},\delta_{i,4},\delta_{i,5}) \hspace{2em} (i=1,2,\cdots,5).
	\end{equation}
	Thus, vertices on the hypercubic lattice are written by $\bm{n}_k=\sum_{i=1}^5 n_{k,i} \bm{e}_i$, where $k$ is the vertex number and $n_{k,i}\in \mathbb{Z}$.
	Here, we introduce the real (perpendicular) space as two-dimensional plane (three-dimensional space) constructed by orthonormal vectors $\bm{v}_1$ and $\bm{v}_2$ ($\bm{v}_3$, $\bm{v}_4$ and $\bm{v}_5$) defined by
	\begin{eqnarray}
	&\bm{v}_1&=\sqrt{\frac{2}{5}}\left(1,\cos \phi,\cos 2\phi,\cos 3\phi,\cos 4\phi\right),\\
	&\bm{v}_2&=\sqrt{\frac{2}{5}}\left(0,\sin \phi,\sin 2\phi,\sin 3\phi,\sin 4\phi\right),\\
	&\bm{v}_3&=\sqrt{\frac{2}{5}}\left(1,\cos 2\phi,\cos 4\phi,\cos \phi,\cos 3\phi\right),\\
	&\bm{v}_4&=\sqrt{\frac{2}{5}}\left(0,\sin 2\phi,\sin 4\phi,\sin \phi,\sin 3\phi\right),\\
	&\bm{v}_5&=\sqrt{\frac{1}{5}}\left(1,1,1,1,1\right)
	\end{eqnarray}
	with $\phi=2\pi/5$. Note that these vectors are orthonormal.
	To project five-dimensional vertices into the real and perpendicular spaces, we use projection matrices,
	\begin{equation}
	\bm{P}_{\mathrm{r}}=\sum_{i=1}^2 \bm{e}_i^{(\mathrm{r})}\otimes \bm{v}_i, \ \bm{P}_{\mathrm{p}}=\sum_{j=1}^3 \bm{e}_j^{(\mathrm{p})}\otimes \bm{v}_{j+2},
	\end{equation}
	where unit vectors in real and perpendicular spaces are given by $\bm{e}_i^{(\mathrm{r})} = (\delta_{i,1},\delta_{i,2})_{\mathrm{r}}$ and $\bm{e}_j^{(\mathrm{p})} = (\delta_{j,1},\delta_{j,2},\delta_{j,3})_{\mathrm{p}}$ for $i=1,2$ and $j=1,2,3$, respectively.
	By using the projection matrices, the real- and perpendicular-space vertices are obtained by $\bm{v}_k^{(\mathrm{r})}=\bm{P}_{\mathrm{r}}\bm{n}_k=(\bm{n}_k\cdot\bm{v}_1,\bm{n}_k\cdot\bm{v}_2)_{\mathrm{r}}$ and $\bm{v}_k^{(\mathrm{p})}=\bm{P}_{\mathrm{p}}\bm{n}_k=(\bm{n}_k\cdot\bm{v}_3,\bm{n}_k\cdot\bm{v}_4,\bm{n}_k\cdot\bm{v}_5)_{\mathrm{p}}$. 
	As confirmed easily, we can find all vertices in a Penrose lattice as the vertices of hypercubic lattice projected into the real space, e.g., five vertices $\bm{n}_k=\bm{e}_k$ for $k=1,2,\cdots,5$ give five apices of pentagon (star) located on the origin.
	
	However, the projected vertices obviously include unwanted vertices for a Penrose lattice.
	To exclude these unwanted vertices, we use a three-dimensional window in the perpendicular space.
	The window is a rhombic icosahedron constructed by five vectors $\bm{d}_i^{(\mathrm{p})} = \bm{P}_{\mathrm{p}}\bm{e}_i$; inner space of the window is given by $\mathscr{W}=\left\{\sum_{i=1}^5r_i\bm{d}_i^{(\mathrm{p})}\ |\ r_i\in [0,1)  \right\}$.
	If a projected vertex into the perpendicular space $\bm{P}_{\mathrm{p}}\bm{n}_k$ is out of the window, we ignore a projected vertex of $\bm{n}_k$ into the real space.
	Through this procedure, we exclude the unwanted vertices for a Penrose lattice~\cite{senechal1996quasicrystals}.
	Note that the allowed vertices $\bm{n}_k$ are classified into four groups by an integer index $Z=\sqrt{5}\bm{n}_k\cdot\bm{v}_5=1,2,3,4$ corresponding to $z$ component of the perpendicular space, i.e., four planes in the perpendicular space.
	Therefore, the four planes restricted in the window include all vertices giving a Penrose lattice.
	
	Next, we move to a Penrose lattice with periodic boundary condition, which corresponds to an approximant of Penrose lattice.
	To obtain the approximant, we use a multigrid method as follows \cite{BABALIEVSKI199027}.
	In this method, we make a Penrose lattice or its approximant in two steps: (i) find a five-dimensional integer vector $\bm{n}(\bm{x})$ as a function of two-dimensional real vector $\bm{x}$, and (ii) make a vertex of the Penrose lattice or its approximant with $\bm{v}^{(\mathrm{r})}(\bm{x})=\bm{P}_{\mathrm{r}}\bm{n}(\bm{x})=\sum_{i=1}^5 n_i(\bm{x}) \bm{d}_i^{(\mathrm{r})}$ where $\bm{d}_i^{(\mathrm{r})} = \bm{P}_{\mathrm{r}}\bm{e}_i$.
	As explained above, if the integer vector $\bm{n}(\bm{x})$ includes all vectors consisting of arbitrary integers $n_{i}\in \mathbb{Z}$, unwanted vertices are also included in a Penrose lattice obtained by the step (ii) with $\bm{n}(\bm{x})$.
	To exclude unwanted vertices, interestingly, we only consider the integer vector given by
	\begin{equation}
	n_i(\bm{x})=\lfloor \bm{x}\cdot \bm{d}_i^{(\mathrm{r})}-\gamma_i\rfloor, \label{MultigraidN}
	\end{equation}
	where the floor function $\lfloor a\rfloor$ denotes the largest integer less than or equal to $a$, and $\gamma_i$ is an arbitrary real number satisfying $\sum_{i=1}^5\gamma_i\in \mathbb{Z}$.
	In this equation, the floor function gives an integer indexing a neighboring vertex of $\bm{x}$, and $\gamma_i$ plays the role of window~\cite{PhysRevB.43.8879}. 
	Therefore, if we search all integer vectors $\bm{n}(\bm{x})$ in a two-dimensional certain finite space $\bm{x}\in \mathscr{S}_r$, we can obtain a finite-size Penrose lattice around the space $\mathscr{S}_r$.
	However, this procedure usually requires a careful searching without dropping any vertices.
	To find the set of integer vectors $\bm{n}(\bm{x})$ efficiently, we use a recursive algorithm proposed in Ref.~\cite{BABALIEVSKI1993370}. 
	
	On the other hand, to approximate the Penrose lattice to a periodic lattice, we substitute in Eq.~(\ref{MultigraidN}) for the quasi unit vectors $\bm{d}_i^{(\mathrm{r})}$ rewritten by,
	\begin{eqnarray}
	&\bm{d}_1^{(\mathrm{r})}&=\sqrt{\frac{2}{5}}(1,0)_{\mathrm{r}},\nonumber\\
	&\bm{d}_2^{(\mathrm{r})}&=\sqrt{\frac{2}{5}}(\cos \phi,\sin \phi)_{\mathrm{r}},\nonumber\\
	&\bm{d}_3^{(\mathrm{r})}&=-\bm{d}_1^{(\mathrm{r})}+\tau^{-1} \bm{d}_2^{(\mathrm{r})},\nonumber\\ &\bm{d}_4^{(\mathrm{r})}&=-\tau^{-1}\left\{\bm{d}_1^{(\mathrm{r})}+\bm{d}_2^{(\mathrm{r})}\right\},\nonumber\\
	&\bm{d}_5^{(\mathrm{r})}&=\tau^{-1}\bm{d}_1^{(\mathrm{r})}-\bm{d}_2^{(\mathrm{r})}.
	\end{eqnarray}
	Here, the golden ratio $\tau=(1+\sqrt{5})/2$ is approximated by a rational number $\tau_n\equiv F_{n+1}/F_{n} \underset{n\to\infty}{\longrightarrow}\tau$, where $F_n$ is the $n$th Fibonacci number.
	With the rational number $\tau_n$, the quasi unit vectors give a large unit cell with translational symmetry.
	Therefore, we obtain an approximant of Penrose lattice as the unit cell including $N_v=4F_{2n+1}+3F_{2n}$ vertices.
	Figure \ref{fig:PBC} represents an approximant with $n=3$, which contains $N_v=76$ vertices as a unit cell.  
	Note that the upper (right) and lower (left) boundaries of this approximant are connected with periodic boundary condition. 
	In this paper, we consider an approximant of Penrose lattice with $n=11$, which results in $N_v=167761$ vertices at most. 
	
	\bibliography{myreference}
\end{document}